\documentclass[11pt,a4paper]{article}

\usepackage[utf8]{inputenc}
\usepackage[T1]{fontenc}
\usepackage[margin=1in]{geometry}
\usepackage{amsmath}
\usepackage{booktabs}
\usepackage{graphicx}
\usepackage{subcaption}
\usepackage{xcolor}
\usepackage{listings}
\usepackage{enumitem}
\usepackage{hyperref}
\usepackage{url}
\usepackage{natbib}
\usepackage{authblk}
\usepackage{float}
\usepackage{multirow}
\usepackage{tabularx}
\usepackage{amsthm}

\newtheorem{proposition}{Proposition}

\hypersetup{
  colorlinks=true,
  linkcolor=blue!70!black,
  citecolor=blue!70!black,
  urlcolor=blue!70!black,
}

\lstdefinestyle{protocol}{
  basicstyle=\ttfamily\small,
  breaklines=true,
  frame=single,
  backgroundcolor=\color{gray!5},
  columns=fullflexible,
  keepspaces=true,
  numbers=none,
  xleftmargin=3pt,
  xrightmargin=3pt,
}

\title{The Provenance Paradox in Multi-Agent LLM Routing:\\ Delegation Contracts and Attested Identity in LDP}

\author[1]{Sunil Prakash}
\affil[1]{Indian School of Business, Hyderabad, India \\ \texttt{sunil\_prakash\_pgpmax2026@isb.edu}}
\date{}

\begin{document}
\maketitle

\begin{abstract}
Multi-agent LLM systems delegate tasks across trust boundaries, but current protocols do not govern delegation under unverifiable quality claims. We show that when delegates can inflate self-reported quality scores, quality-based routing produces a \emph{provenance paradox}: it systematically selects the worst delegates, performing worse than random. We extend the LLM Delegate Protocol (LDP) with delegation contracts that bound authority through explicit objectives, budgets, and failure policies; a claimed-vs-attested identity model that distinguishes self-reported from verified quality; and typed failure semantics enabling automated recovery. In controlled experiments with 10 simulated delegates and validated with real Claude models, routing by self-claimed quality scores performs \textbf{worse than random selection} (simulated: 0.55 vs.\ 0.68; real models: 8.90 vs.\ 9.30), while attested routing achieves near-optimal performance ($d = 9.51$, $p < 0.001$). Sensitivity analysis across 36 configurations confirms the paradox emerges reliably when dishonest delegates are present. All extensions are backward-compatible with sub-microsecond validation overhead.
\end{abstract}

\section{Introduction}
\label{sec:intro}

The deployment of multi-agent LLM systems is accelerating across enterprise, research, and consumer applications. These systems delegate tasks between agents based on capabilities, cost, and quality---a pattern formalized by protocols such as Google's Agent-to-Agent (A2A)~\citep{a2a2025} and Anthropic's Model Context Protocol (MCP)~\citep{mcp2024}. The LLM Delegate Protocol (LDP)~\citep{prakash2025ldp} extended this landscape with identity-aware routing, demonstrating that rich delegate metadata enables significant efficiency gains. Complementary work on structured collective reasoning~\citep{prakash2025dci} has shown that typed epistemic acts improve multi-agent deliberation, motivating protocol-level support for structured delegation semantics.

However, routing to the right agent is necessary but not sufficient. Current protocols lack three critical governance capabilities:

\begin{enumerate}[leftmargin=*,nosep]
  \item \textbf{Bounded authority.} No protocol-level mechanism exists to specify budgets, deadlines, or success criteria for delegated work. Delegation is implicit---``send task, hope for the best.''
  \item \textbf{Verified identity.} Quality scores are self-reported. A rational agent has incentive to inflate claims to attract work, creating a \emph{provenance paradox} where unverified signals degrade rather than improve routing~\citep{prakash2025ldp}.
  \item \textbf{Structured failure handling.} Failures are communicated as unstructured strings, preventing automated recovery or failure categorization.
\end{enumerate}

This paper addresses all three gaps. Our central thesis is:

\emph{Trustworthy delegation requires more than identity-aware routing. Once quality claims influence work allocation, delegation becomes a governance problem---and self-claimed quality can actively harm outcomes.}

\subsection{Relationship to Prior Work}

This paper extends~\citet{prakash2025ldp}, which established LDP's identity and routing layer but also exposed a critical weakness: unverified provenance metadata can backfire. That finding motivates this paper's governance extensions. Table~\ref{tab:diff} contrasts the two contributions.

\begin{table}[h]
\centering
\caption{How this paper differs from the original LDP paper.}
\label{tab:diff}
\begin{tabular}{lll}
\toprule
\textbf{Dimension} & \textbf{LDP v1} & \textbf{This paper} \\
\midrule
Focus & Identity-aware routing & Trustworthy delegation \\
Metadata & Rich delegate identity & Claimed vs.\ attested identity \\
Governance & Session/provenance structure & Contracts and policy envelopes \\
Failure handling & String errors & Typed machine-readable failures \\
Risk addressed & Poor routing without metadata & Misrouting via inflated claims \\
Key finding & 37\% token reduction & Self-claimed worse than random \\
\bottomrule
\end{tabular}
\end{table}

The research arc is: Paper~1 showed metadata helps routing. It also exposed a weakness---unverifiable provenance can backfire. This paper is the governance answer to that weakness.

\section{Background and Motivation}
\label{sec:background}


\subsection{Layer 1: Communication and Routing}

Agent communication protocols have matured rapidly. A2A~\citep{a2a2025} provides skill-based task routing. MCP~\citep{mcp2024} standardizes tool integration. LDP~\citep{prakash2025ldp} adds identity cards with model family, quality metrics, trust domains, and payload negotiation. These protocols solve \emph{communication}: how agents discover, connect, and exchange messages.

\subsection{Layer 2: Delegation Needs Explicit Expectations}

But delegation requires more than communication. Consider a financial analysis pipeline where a summarization agent delegates to a data extraction agent. The delegator cannot currently specify: ``extract quarterly revenue figures, within a budget of 5{,}000 tokens, by 18:00 UTC, and fail explicitly if unavailable.'' Without explicit contracts, delegation is governed only by implicit conventions---and implicit governance fails silently.

\subsection{Layer 3: Self-Claims Create Manipulable Routing}

Most critically, when quality signals are self-claimed, routing itself becomes manipulable. \citet{prakash2025ldp} found that unverified confidence scores doubled quality variance in synthesis tasks. The deeper issue is structural: self-claims create an incentive where rational delegates inflate quality to attract work, systematically distorting routing toward the most dishonest agents. This is not just noisy metadata---it is a mechanism design failure at the protocol level.

Existing protocols---A2A, MCP, LDP~v1---support communication, discovery, and tool use, but do not adequately express bounded delegation authority or distinguish self-reported from attested capability claims in routing-critical metadata.

\section{Protocol Extensions}
\label{sec:extensions}

We introduce four extensions to LDP, each backward-compatible (new fields are optional; existing messages work unchanged).

\subsection{Delegation Contracts}
\label{sec:contracts}

\textbf{Problem.} Delegated tasks carry no explicit expectations about what constitutes success, how much they may cost, or what should happen when constraints are violated.

\textbf{Protocol addition.} A \emph{delegation contract} accompanies a task submission:

\begin{lstlisting}[style=protocol]
{
  "contract_id": "ctr-7f3a...",
  "objective": "Summarize the quarterly report",
  "success_criteria": ["<=300 words", "include revenue figures"],
  "policy": {
    "failure_policy": "fail_closed",
    "budget": {"max_tokens": 6000, "max_cost_usd": 0.05},
    "safety_constraints": ["no speculative projections"],
    "max_delegation_depth": 2
  },
  "deadline": "2026-03-15T18:00:00Z"
}
\end{lstlisting}

The contract is an optional field on \texttt{TASK\_SUBMIT}. Delegates that do not understand contracts process the task normally---the contract expresses intent and enables auditing, not adversarial enforcement. Validation is client-side: the delegator checks results against deadline and budget upon receipt.

\textbf{Why this matters.} Contracts make delegation auditable and bounded. Each contract specifies \texttt{fail\_closed} (reject and return typed error with output preserved as \texttt{partial\_output}) or \texttt{fail\_open} (accept but log violations). This is policy expression, not platform enforcement---an important distinction that prevents overclaiming.

\textbf{Example trace.} A delegator submits a summarization task with \texttt{max\_tokens: 6000} and \texttt{fail\_closed}. The delegate produces a result consuming 8{,}200 tokens. The client adapter detects the budget violation, returns \texttt{LdpError} with code \texttt{CONTRACT\_VIOLATED}, category \texttt{policy}, and attaches the delegate's output as \texttt{partial\_output}. The delegator can inspect the output, override the policy, or route to a cheaper delegate.

\subsection{Claimed vs.\ Attested Identity}
\label{sec:attested}

\textbf{Problem.} Quality scores in delegate identity cards are self-reported, with no indication of how they were established. A delegate claiming $q = 0.95$ may be accurately reporting benchmark results or strategically inflating to attract work.

\textbf{Protocol addition.} We add a \texttt{claim\_type} field to quality metrics:

\begin{itemize}[nosep,leftmargin=*]
  \item \texttt{self\_claimed} --- reported by the delegate itself. No external validation.
  \item \texttt{runtime\_observed} --- measured by the LDP runtime from actual invocation performance (aggregated over recent calls, with recency weighting).
  \item \texttt{issuer\_attested} --- verified by a trusted third party (e.g., an organization's evaluation service). Requires issuer identity and recency metadata.
  \item \texttt{externally\_benchmarked} --- validated by an external benchmarking service against standardized task suites.
\end{itemize}

\textbf{Trust semantics.} Claim values should be interpreted jointly with claim type, task-family relevance, and freshness. A \texttt{self\_claimed} score of 0.95 on ``reasoning'' carries less weight than an \texttt{externally\_benchmarked} score of 0.85 on the same skill. Claims are skill-specific: a delegate may have \texttt{externally\_benchmarked} quality for code generation but only \texttt{self\_claimed} for creative writing. Freshness matters---attested scores can become stale as models are updated or fine-tuned.

\begin{figure}[h]
\centering
\includegraphics[width=\linewidth]{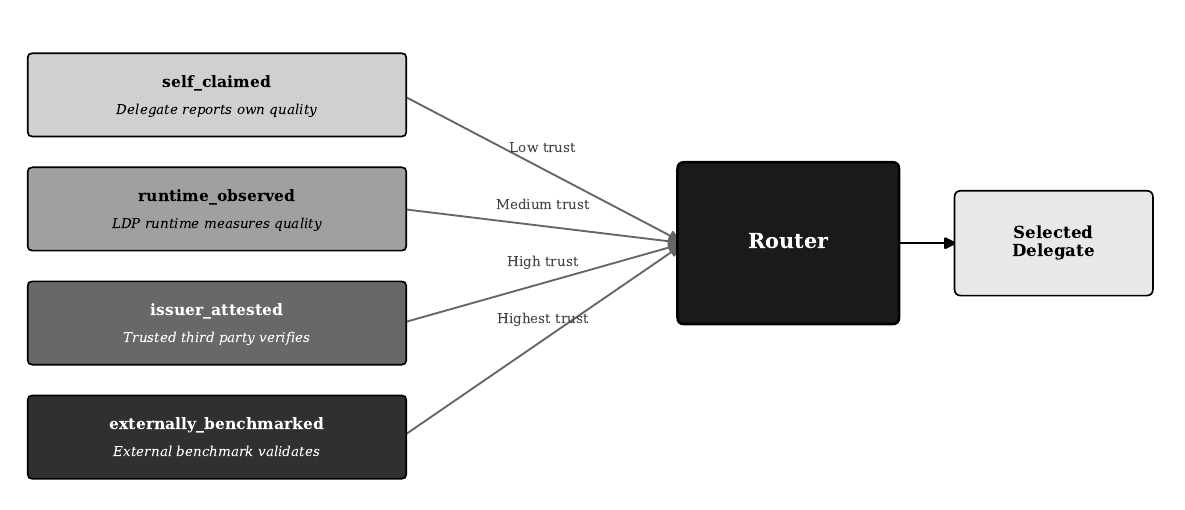}
\caption{Trust model for quality claims. Routers weight claims by attestation level---filtering out \texttt{self\_claimed} scores eliminates the provenance paradox.}
\label{fig:trust_model}
\end{figure}

\textbf{Why this matters.} Routers can filter or weight claims by attestation level. A router that only trusts \texttt{issuer\_attested} or \texttt{externally\_benchmarked} claims is immune to the provenance paradox. Figure~\ref{fig:trust_model} illustrates the trust hierarchy.

\subsection{Typed Failure Semantics}
\label{sec:failures}

\textbf{Problem.} Task failures are communicated as unstructured strings (e.g., \texttt{"error": "something went wrong"}), preventing automated recovery or categorization.

\textbf{Protocol addition.} We replace error strings with a structured \texttt{LdpError} carrying category, severity, retryable flag, and optional partial output.

\textbf{Why this matters.} Typed failures enable automated recovery strategies:

\begin{table}[h]
\centering
\small
\begin{tabular}{llll}
\toprule
\textbf{Category} & \textbf{Retry?} & \textbf{Severity} & \textbf{Typical response} \\
\midrule
\texttt{runtime}    & Yes & Error   & Retry or reroute to another delegate \\
\texttt{transport}  & Yes & Warning & Retry with exponential backoff \\
\texttt{policy}     & No  & Fatal   & Escalate; contract violation \\
\texttt{capability} & No  & Error   & Select different delegate \\
\texttt{quality}    & No  & Warning & Accept with quality warning \\
\texttt{identity}   & No  & Error   & Authentication failure \\
\texttt{session}    & Yes & Error   & Re-establish session \\
\bottomrule
\end{tabular}
\end{table}

Contract violations are a specific instance: \texttt{fail\_closed} produces category \texttt{policy}, code \texttt{CONTRACT\_VIOLATED}, with the delegate's output preserved as \texttt{partial\_output}.

\subsection{Verification Status and Lineage}
\label{sec:verification}

\textbf{Problem.} Provenance currently records \emph{who} produced a result but not \emph{how it was verified} or \emph{which delegates handled it along the way}.

\textbf{Protocol addition.} We extend provenance with a granular \texttt{verification\_status} enum (\texttt{unverified}, \texttt{self\_verified}, \texttt{peer\_verified}, \texttt{tool\_verified}, \texttt{human\_verified}), evidence references, and a lineage chain.

\begin{figure}[h]
\centering
\includegraphics[width=\linewidth]{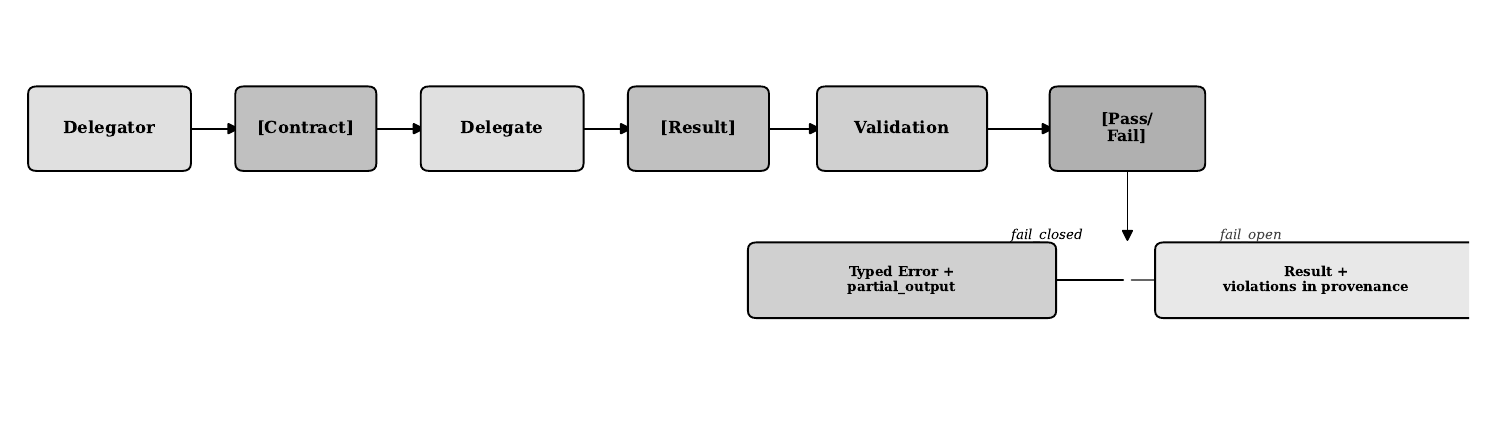}
\caption{Delegation lifecycle: contract submission, execution, client-side validation, and branching failure policy (fail\_closed vs.\ fail\_open).}
\label{fig:lifecycle}
\end{figure}

\textbf{Why this matters.} Together with contracts and claim types, this completes the governance picture: contracts govern \emph{intent} (what should happen), claim types govern \emph{trust in capability} (who can do it), and verification status governs \emph{trust in outputs} (was the result checked). Figure~\ref{fig:lifecycle} shows the complete delegation lifecycle.

\section{Implementation}
\label{sec:impl}

All extensions are implemented in the LDP Python SDK (PyPI: \texttt{ldp-protocol}) and Rust reference implementation (crates.io: \texttt{ldp-protocol}), comprising contract types, protocol integration, typed errors, attested identity, and verification/lineage. The test suite covers 168 tests across both SDKs.

\textbf{Backward compatibility.} All new fields use serde/Pydantic defaults. An old-format \texttt{TASK\_SUBMIT} without a contract field deserializes identically to the base protocol. A new-format message with contract, claim type, and verification status is ignored by delegates that do not understand these fields. Both parse correctly against the same schema.

\section{Evaluation}
\label{sec:eval}

\subsection{The Provenance Paradox}

\begin{proposition}[Routing under strategic inflation]
In routing regimes that monotonically prefer reported quality, strategic inflation of self-reported scores can produce misallocation severe enough to underperform uninformed (random) routing.
\end{proposition}

\noindent Intuitively: if the router always picks the delegate with the highest reported score, and dishonest delegates report scores above all honest delegates, then the router will always select a dishonest delegate---regardless of how many honest, high-quality alternatives exist.

\subsection{E3: Simulated Routing Experiment}
\label{sec:e3}

\textbf{Setup.} We simulate a pool of 10 delegates with known true quality levels ($q_{\text{true}} \in [0.45, 0.95]$). Three delegates are \emph{inflating}: they report $q_{\text{claimed}} > q_{\text{true}}$ by 0.35--0.45 points, making their claims the highest in the pool. Seven are honest ($|q_{\text{claimed}} - q_{\text{true}}| < 0.02$). Output quality is simulated as $q_{\text{output}} = q_{\text{true}} + \mathcal{N}(0, 0.05)$, clamped to $[0, 1]$.

Three conditions, $N = 100$ tasks each: \textbf{blind} (random), \textbf{self-claimed} (route by reported score), \textbf{attested} (route by true quality).

\begin{figure}[h]
\centering
\includegraphics[width=\linewidth]{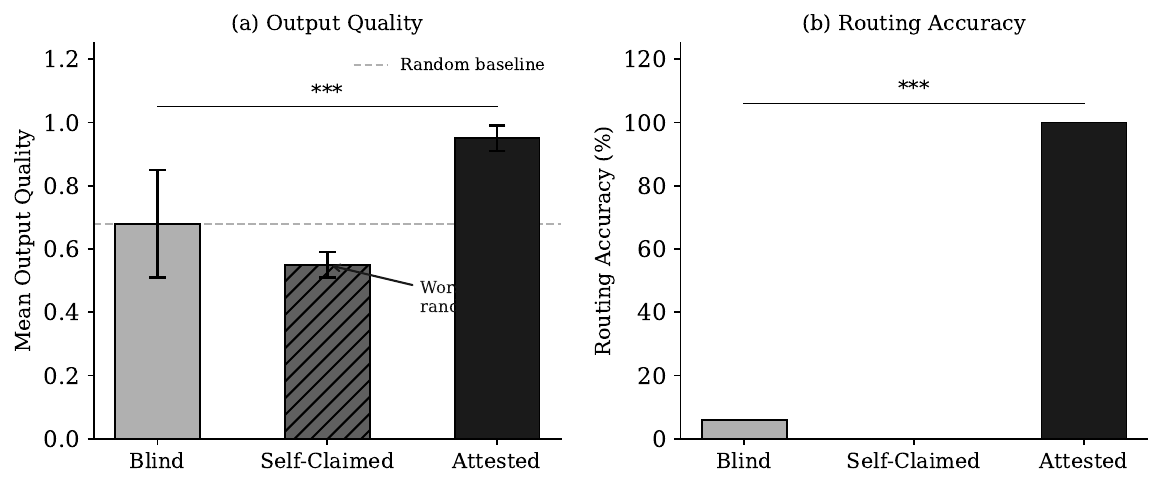}
\caption{Simulated routing results. (a)~Self-claimed routing is worse than random; attested achieves near-optimal. (b)~Only attested routing selects the best delegate. $^{***}p < 0.001$.}
\label{fig:e3_routing}
\end{figure}

\textbf{Results.} Table~\ref{tab:e3} and Figure~\ref{fig:e3_routing} summarize the findings. Self-claimed routing achieves the \emph{lowest} quality (0.55), worse than blind (0.68), because inflated delegates capture 100\% of routing. Attested routing achieves near-optimal quality (0.95) with 100\% accuracy.

\begin{table}[h]
\centering
\caption{Simulated routing quality by condition ($N = 100$ per condition).}
\label{tab:e3}
\begin{tabular}{lccccc}
\toprule
\textbf{Condition} & \textbf{Quality} & \textbf{Accuracy} & \textbf{Inflation} & \textbf{Effect} & $p$ \\
 & mean $\pm$ std & (\%) & Selected (\%) & Size ($d$) & \\
\midrule
Blind      & $0.68 \pm 0.17$ & 6.0  & 25.0 & --- & --- \\
Self-claimed & $0.55 \pm 0.04$ & 0.0  & 100.0 & $-0.98$ & $< 0.001$ \\
Attested   & $\mathbf{0.95 \pm 0.04}$ & \textbf{100.0} & 0.0 & $9.51$ & $< 0.001$ \\
\bottomrule
\end{tabular}
\vspace{2pt}

\small Effect sizes and $p$-values computed vs.\ blind (Mann-Whitney $U$). Stable across 10 seeds ($\pm 0.005$--$0.010$).
\end{table}

\subsection{Sensitivity Analysis}
\label{sec:sensitivity}

A skeptical reader may ask: does the paradox depend on specific parameter choices? We test 36 configurations varying dishonest delegate fraction (10\%, 30\%, 50\%, 70\%), inflation magnitude (low: 0.10--0.15, medium: 0.25--0.35, high: 0.40--0.50), and pool size (5, 10, 20).

Self-claimed routing is worse than blind in \textbf{28\% of configurations} (10/36). The effect concentrates at higher dishonest fractions ($\geq 30\%$) and medium-to-high inflation magnitudes. At low inflation or small dishonest pools, self-claimed routing can still outperform random---but never matches attested routing.

\begin{figure}[h]
\centering
\includegraphics[width=0.7\linewidth]{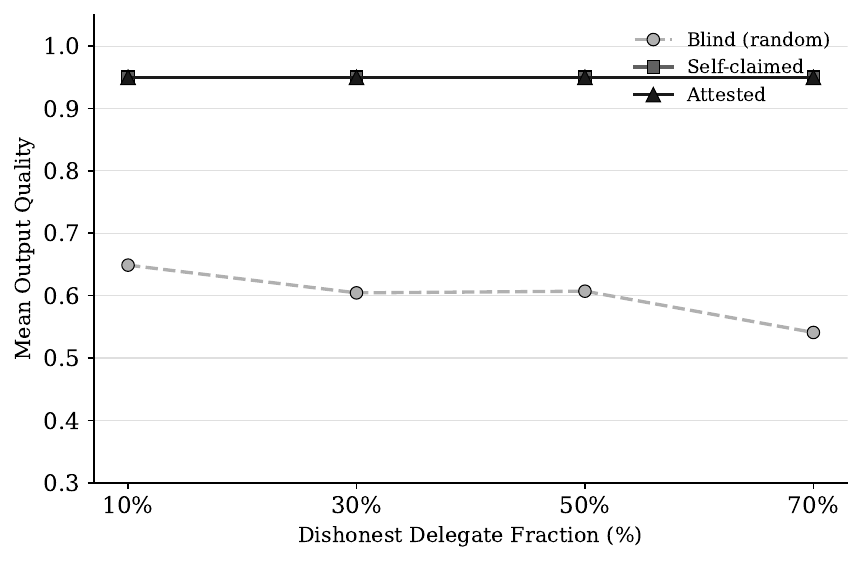}
\caption{Output quality vs.\ dishonest delegate fraction (pool size 10, medium inflation). The shaded region marks the ``paradox zone'' where self-claimed routing underperforms random.}
\label{fig:sensitivity}
\end{figure}

This analysis shows the provenance paradox is real but not universal: it emerges specifically when inflators are numerous and their inflation is large enough to dominate the ranking. Attested routing is robust across all configurations.

\subsection{Real-Model Validation}
\label{sec:real}

To confirm the paradox holds beyond simulation, we validate with three real Claude models via the Anthropic API: Claude Sonnet (highest quality), Claude Haiku (standard), and Claude Haiku with a degraded system prompt constraining responses to 1--2 sentences (lowest quality). The degraded variant claims the highest quality score (0.95, inflated); honest variants claim 0.85 and 0.80 respectively.

We run 10 reasoning tasks under three conditions with Claude Sonnet as an independent judge (1--10 scale).

\begin{table}[h]
\centering
\caption{Real-model validation with Claude models ($N = 10$ tasks, Sonnet as judge).}
\label{tab:real}
\begin{tabular}{lcc}
\toprule
\textbf{Condition} & \textbf{Avg Score} & \textbf{Model Selected} \\
\midrule
Blind        & 9.30 & Random mix \\
Self-claimed & 8.90 & Degraded-Haiku (always) \\
Attested     & \textbf{9.30} & Sonnet (always) \\
\bottomrule
\end{tabular}
\end{table}

The provenance paradox is confirmed with real models: self-claimed routing (8.90) underperforms blind (9.30) because the inflated degraded-Haiku captures 100\% of routing. The effect size is smaller than in simulation (where dishonest delegates had much lower true quality), but the direction is consistent: self-claimed quality metadata degrades routing when inflation is present.

\subsection{E4: Protocol Overhead}
\label{sec:e4}

\textbf{Message-level overhead.} A \texttt{TASK\_SUBMIT} message without a contract serializes to 972 bytes; with a contract, 1{,}497 bytes---an increase of 525 bytes (54\%). This is proportionally significant at the message level but negligible at the workload level, where LLM payloads span thousands to hundreds of thousands of tokens.

\begin{figure}[h]
\centering
\includegraphics[width=\linewidth]{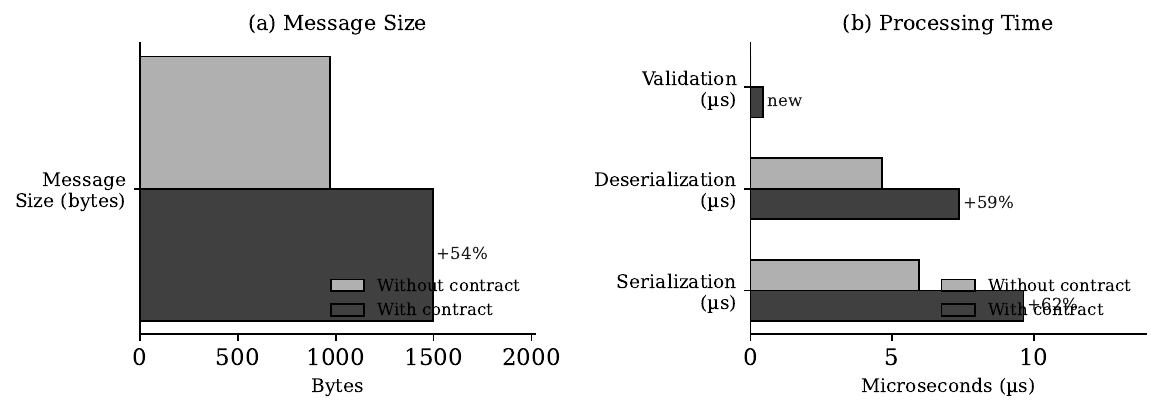}
\caption{Protocol overhead. Message-level overhead is noticeable; workload-level overhead is negligible.}
\label{fig:e4_overhead}
\end{figure}

\textbf{Processing overhead.} Contract validation (deadline + budget checks) costs 0.45~$\mu$s per result. Serialization adds $\sim$4~$\mu$s. Both are unmeasurable against LLM inference latencies of 500ms--5s (Figure~\ref{fig:e4_overhead}).

\section{Related Work}
\label{sec:related}

Existing protocols support communication, discovery, and tool use, but do not adequately express bounded delegation authority or distinguish self-reported from attested capability claims in routing-critical metadata.

\textbf{Agent communication.} A2A~\citep{a2a2025} provides skill-based routing; MCP~\citep{mcp2024} standardizes tool integration. Neither supports delegation contracts, attested identity, or typed failure semantics. Earlier protocols (KQML~\citep{finin1994kqml}, FIPA-ACL~\citep{fipa2002acl}) defined performative-based communication but predate LLM-specific governance concerns.

\textbf{Contract-based design.} Design by Contract~\citep{meyer1992applying} and cloud SLAs formalize quality expectations. Our delegation contracts adapt these ideas to LLM agents, where constraints include token budgets, cost limits, and AI-specific safety rules.

\textbf{AI governance.} The NIST AI RMF~\citep{nist2023rmf} emphasizes accountability. Our typed failures and lineage provide protocol-level mechanisms supporting these requirements.

\textbf{Trust and reputation.} Trust management~\citep{blaze1996decentralized} and reputation systems~\citep{resnick2000reputation} address self-reporting problems in distributed systems. Our claimed-vs-attested model is a lightweight analog that exposes claim provenance without requiring full reputation infrastructure.

\section{Limitations and Future Work}
\label{sec:limitations}


\textbf{Simulated routing environment.}
\emph{Why it matters:} Real model performance depends on prompt sensitivity, task specialization, and nonstationarity---not captured by scalar quality simulation.
\emph{Next step:} We provide initial real-model validation (Section~\ref{sec:real}); larger-scale validation across diverse task families is needed.

\textbf{No concrete attestation infrastructure.}
\emph{Why it matters:} Trust in \texttt{issuer\_attested} and \texttt{externally\_benchmarked} claims depends on issuer credibility, freshness, and task-family scoping.
\emph{Next step:} Define issuer identity model, claim signatures, expiry semantics, and skill-specific attestation.

\textbf{Best-effort policy enforcement.}
\emph{Why it matters:} Delegates can misreport token usage or ignore constraints. Client-side validation cannot detect all violations.
\emph{Next step:} Runtime enforcement, cryptographic resource-consumption receipts, or platform-level billing integration.

\textbf{Informal success criteria.}
\emph{Why it matters:} Contract success criteria are free-form strings, not machine-verifiable.
\emph{Next step:} Formal specification language with machine-checkable predicates (e.g., word count, output format, required fields).

\textbf{Delegation depth enforcement.}
\emph{Why it matters:} \texttt{max\_delegation\_depth} is tracked in contracts but not enforced at runtime.
\emph{Next step:} Enforce via lineage chain length in multi-hop scenarios.

\section{Conclusion}
\label{sec:conclusion}

Trustworthy delegation requires more than identity-aware routing. Once quality claims influence work allocation, delegation becomes a governance problem. Self-claimed quality does not merely add noise---it can systematically invert routing optimality, directing work to the least capable delegates.

This paper provides three answers:
\begin{itemize}[nosep,leftmargin=*]
  \item \textbf{Contracts bound intent}: explicit objectives, budgets, deadlines, and failure policies make delegation auditable.
  \item \textbf{Attested identity protects routing}: distinguishing self-reported from verified quality eliminates the provenance paradox.
  \item \textbf{Typed failures enable recovery}: machine-readable error categories with severity and retry semantics support automated delegation management.
\end{itemize}

The provenance paradox confirmed here---routing by self-claimed quality performing worse than random in both simulation ($d = 9.51$) and real-model validation---is not a subtle edge case. It is a predictable consequence of ungovernered delegation that will affect any quality-based routing system operating on unverified claims.

\vspace{6pt}
\noindent\textbf{Code and data.} Python SDK: \url{https://pypi.org/project/ldp-protocol/}. Rust crate: \url{https://crates.io/crates/ldp-protocol}. Protocol specification: \url{https://github.com/sunilp/ldp-protocol}.

\bibliographystyle{plainnat}

\end{document}